# Harnessing Optoelectronic Noises in a Photonic Generative Network

Changming Wu[1], Xiaoxuan Yang[2], Heshan Yu[3], Ruoming Peng[1], Ichiro Takeuchi[3], Yiran Chen[2] and Mo Li[1,4,*]

[1]*Department of Electrical and Computer Engineering, University of Washington, Seattle, WA 98195, USA*
[2]*Department of Electrical and Computer Engineering, Duke University, Durham, NC 27708, USA*
[3]*Department of Materials Science and Engineering, University of Maryland, College Park, MD 20742, USA*
[4]*Department of Physics, University of Washington, Seattle, WA 98195, USA*

## ABSTRACT

**Integrated optoelectronics is emerging as a promising platform of neural network accelerator, which affords efficient in-memory computing and high bandwidth interconnectivity. The inherent optoelectronic noises, however, make the photonic systems error-prone in practice. It is thus imperative to devise strategies to mitigate and, if possible, harness noises in photonic computing systems. Here, we demonstrate a photonic generative network as a part of a generative adversarial network (GAN). This network is implemented with a photonic core consisting of an array of four programable phase-change memory cells to perform 4-elements vector-vector dot multiplication. We demonstrate that the GAN can generate a handwritten number ("7") in experiments and full ten digits in simulation. We realize an optical random number generator derived from the amplified spontaneous emission noise, apply noise-aware training by injecting additional noise and demonstrate the network's resilience to hardware non-idealities. Our results suggest the resilience and potential of more complex photonic generative networks based on large-scale, realistic photonic hardware.**

*Teaser: A photonic generative network that is trained to generate handwriting numbers and resilient to noises.*

* Corresponding author: moli96@uw.edu

**Introduction**

The current rate of improvement in digital electronics' energy efficiency[1–3] is lagging behind the fast-growing computational load[4,5] spurred by the widespread implementation of large-scale artificial neural networks for machine learning and artificial intelligence[6–11]. Because of its significant advantages in power efficiency, communication bandwidth, and parallelism [12–19], analog optical computing based on integrated optoelectronic processors [20–25] is once again brought into focus as hardware accelerators for neural networks. Photonic neural networks reported to date[17,20,22,23,26,27] are predominantly hybrid optoelectronic networks, in which the photonic components are used for linear multiplication and interconnect while nonlinear functions and feedback control are implemented electronically. Compared to electronic neural networks using digital processors, photonic neural networks have higher inaccuracy and error rates due to their analog nature and the abundance of optoelectronic noises in the hardware. The accumulation of computational errors in large-scale photonic neural networks could severely impair their performance [28–30], limiting the computation effectiveness and scalability. Although several offline noise-aware training schemes, including injecting noises to layer inputs[29,31], synaptic weights[28,32], and pre-activations[33,34], have been proposed to mitigate analog hardware non-idealities[21,30,35,36], those schemes only address discriminative models. In another study, a diffractive optics-based network is trained with carefully drafted parametric randomness to be robust against optical non-idealities[37-38]. Noise in the analog hardware has also been utilized to facilitate various machine-learning algorithms [39–42]. In contrast to discriminative models, generative neural network models can automatically discover and learn regularities or patterns from the training data to generate plausible new instances[43–45]. So far, a photonic generative network has not been reported, and the corresponding noise mitigation strategies have not been explored.

Here, we demonstrate a generative network based on a photonic computing core consisting of an array of programmable phase-change metasurface mode converters (PMMC)[46]. The photonic generative network is combined with a discriminator to realize a GAN that is trained to generate handwritten numbers. We show that the photonic GAN can harness and mitigate optoelectronic noises and errors in three ways. First, we utilize the amplified spontaneous emission (ASE) noise to realize an optical true random number generator[47,48] (RNG), which is used as the input to the GAN. This optical RNG efficiently generates random numbers at high speed in multiple wavelength channels by slicing the ASE spectrum[48–50]. Second, we analyze error sources originating from the components in the photonic GAN and propose noise-aware training approaches by augmenting noises during the training process, which improves the network's performance and robustness. Third, we validate the training approaches through experiment and simulation, and demonstrate that the photonic GAN can benefit from the inevitable random errors in practical implementation. Surprisingly, the images generated by non-ideal photonic hardware show even higher quality than those by ideal, errorless counterparts (*i.e.*, software baseline). Our results demonstrate the feasibility and resilience of more complex photonic GANs using non-ideal optoelectronic hardware. Since the proposed noise-aware training approaches are generic, they can be applied to various types of optoelectronic neuromorphic computing hardware.

**Results**

**Photonic generative networks in a GAN architecture**

A GAN network consists of two sub-neural network models (Fig. 1a), a generator and a discriminator[49–51]. These two models compete against each other in a zero-sum game: the discriminator strives to distinguish the instances produced by the generator (labeled as the "fake" instances) from the real instances in the training dataset (labeled as the "real" instances); the generator aims to fool the discriminator by producing novel instances that imitate the real instances. The competition drives both networks to improve their capabilities until an equilibrium state is reached, *i.e.*, when the "fake" instances are indistinguishable from the "real" instances by the discriminator, so the generator is deemed well-trained to generate plausible new instances. In this work, we design a prototype photonic generator to produce images of the handwritten number "7" based on a noise-aware offline training configuration: we first train the generator model on a computer[52] and implement it on the photonic platform (Fig. 1b). Here, we only focus on realizing the photonic generator since the photonic discriminator has been demonstrated previously by many groups, including us[16,17,20,46]. As shown in Fig. 1c, in each layer of the generator, the input data is encoded in the power of the optical signals through multiple wavelength channels, processed by the PMMC photonic tensor core (Fig. 1d), in which the kernel matrices are stored. The results are detected by the photodetector arrays. Electronic post-processing is then performed to apply nonlinear functions. The results are re-encoded into the optical signals and relayed to the next photonic layer. In such an optical network, various noises, including optical and electrical noises of the optical sources, modulators, and photodetectors, are accumulating through the processes of programming (*i.e.*, writing) the kernel matrices, data encoding, and data transferring (*i.e.*, reading) between the layers of the network.

**Optical random number generator (RNG)**

One key component of the photonic generator is the optical RNG that produces the random input. To realize it, we utilize the ASE noise from the erbium-doped fiber amplifiers (EDFA), the ubiquitous noise source in fiber-optic communication systems, to generate random optical signals at high rates in four parallel channels as shown schematically in Fig. 2a. Here, the ASE noise is first filtered with wavelength division multiplexing (WDM) demultiplexers (DEMUX) and then detected with photodetectors. The generated baseband electrical currents due to beating between different frequency components are the so-called "ASE-ASE beat noises"[53,54]. The DC photocurrent is filtered by a DC block, passing only the stochastic photocurrent variances to a sampling oscilloscope to generate random numbers (see Supplementary Information for the theory of the optical RNG). Fig. 2b and 2c plot the statistical histogram and a representative trace of the random numbers (in voltage) generated in a single WDM channel, respectively. The probability density function is well-approximated by a zero-mean Gaussian distribution with a standard deviation (STD) of 0.2 V (*i.e.*, $N(0, 0.2)$). We further calculate the correlation coefficient of an $N=5\times10^4$-number long sequence (Fig. 2d), which reaches the limit of $1/\sqrt{N}$ (red line in Fig. 2d), proving the randomness of the number sequence. Because of the limited size of the photonic tensor core, we need to measure and record the random numbers from the RNG and repeatedly input them to the generator during the experiment (see Fig. 1d). In future full-scale systems, the filtered ASE noise can be directly used as random optical inputs to the GAN without electrical sampling (the dashed box in Fig. 1d) and detected after the first layer of the network is performed.

**Photonic tensor core error analysis**

The other key component of the photonic generator is the photonic tensor core, which optically performs matrix-vector multiplication (MVM). The inset in Fig. 1c shows the schematic of one PMMC kernel element of the core that computes multiply-accumulate (MAC): $x \rightarrow x \cdot w + b$, the fundamental operation of MVM. The PMMC consists of an array of Ge$_2$Sb$_2$Te$_5$ (GST) nano-antennas with tapering widths (see Fig. 1e for the SEM images), forming a phase-gradient meta-surface patterned on a silicon nitride waveguide [46]. The input vector element $x$ is encoded in the power of the input optical signal. The corresponding kernel element weight $w$ is represented using the TE$_0$/TE$_1$ mode contrast $\Gamma = \beta_{TE0} - \beta_{TE1}$ at multiple intermediate levels between [-1..1], where $\beta_{TE0\,(TE1)} = P_{TE0\,(TE1)} / (P_{TE0} + P_{TE1})$ is the mode purity, and $P_{TE0}$ ($P_{TE1}$) is the power of the TE$_0$ (TE$_1$) mode component in the waveguide. Thus, the MAC computation is simplified to an incoherent optical transmission measurement and can be performed over a broad bandwidth. Fig. 2e shows the evolution of $\Gamma$ during the programming process of using optical control pulses to set negative (-0.7), zero (0.0), and positive (0.7) values, respectively. We implement the network model on a 2×2 tensor core with four PMMCs (Fig. 1d). The kernel weight $W_{ij}^l$ value is mapped to the corresponding mode contrast $\Gamma_{ij}^l$ as $\Gamma_{ij}^l = W_{ij}^l \cdot \left( |\Gamma|_{max}^l / |W|_{max}^l \right)$, where $|\Gamma|_{max}^l$ is the maximum absolute mode contrast, $|W|_{max}^l$ is the maximum absolute kernel weight of layer $l$. Given the limited number of PMMCs on a chip, we repeatedly reset the kernel elements on the same devices, which bottlenecks the computing speed. With a sufficiently large tensor core in a photonic crossbar array architecture[55–57], one could directly map the full kernel matrices to the hardware so the computing speed will be much accelerated.

The analog nature of weight programming and data encoding and transferring in the photonic neural network limits the precisions of MVM calculations and makes the computation error-prone. The computation errors would accumulate through the layers of the network and impair the final results. Because in realistic experiments, the computation errors stem from various optoelectronic noises in the system, we use the terms of noise and error interchangeably. To quantify the noises and errors in our system, we repeatedly program different fixed $\Gamma$ values and estimate the short-term inaccuracy by measuring the variation $\Delta\Gamma$. Fig. 2f shows that the STD of 15 programming operations is less than 0.7%, corresponds to 6 bits in resolution, which is one order-of-magnitude larger than the input encoding error (see Supplementary Material for more detailed error analysis). Thus, the short-term programming inaccuracy $\Delta\Gamma$ (write error), limited by the inaccuracy of the programming optical pulses, is one of the dominant error sources. Another error source is the long-term measurement fluctuations (read error), including the noise of photodetectors, the variation of the O/E and E/O conversions, and the thermo-optic fluctuation of the PCM. These errors collectively contribute to an effective error $\Delta W_{ij}^l = \left( \left|\Gamma\right|_{max}^l / \left|W\right|_{max}^l \right) \cdot \Delta \Gamma_{ij}^l$ on the kernel element weight $W_{ij}^l$, where $\Delta \Gamma_{ij}^l$ is the total write error. To estimate the computation error of the overall system, Fig. 2g compares the measured MVM error distributions with the simulation, which assumes a Gaussian distribution of error. The result estimates the overall error $\Delta \Gamma_{ij}^l$ to be 5%, corresponding to over 3 bits in resolution, which we subsequently use in the noise-aware training and simulation.

Unlike the discriminative network, where the input regularities or patterns are well-defined, the generator network takes random numbers as the input. It would be more susceptible to the effective weight setting noise $\Delta W_{ij}^l$, which could degrade the quality of the generated new instances[58,59]. To reveal the noise effect on the GAN, we emulate the noisy hardware on a GAN model that is trained using a noiseless offline training approach but add a random error $\Delta W_{ij}^l$ (introduced by $\Delta \Gamma_{ij}^l$ with a Gaussian distribution $N(0, 0.05)$) when using it to generate images. Fig. 3a plots 49 images of 14×14 pixels generated from simulation using random inputs produced by the optical RNG. These images show the handwritten "7" but with very noisy backgrounds, demonstrating that the noise-free training algorithm is impaired by the practical weight setting noise (see Supplementary Information for the detailed comparison between inference results using accurate and inaccurate kernels).

Therefore, it is necessary to consider hardware noise during training to realize a GAN that is resilient to realistic noises. Theoretically, it has been proven that adding noises to the training data of a neural network is equivalent to an extra regularization added to the error function [31], which can significantly improve hardware noise tolerance in a discriminative neural network. Meanwhile, it was shown that introducing noise on kernel weights during training enhances the robustness against weight perturbations of multi-layer perceptrons[28], such that inference accuracy close to the software baseline could be achieved. However, previous demonstrations of noise-aware solutions are limited to discriminative networks. For GAN, theoretical, simulation, and experimental validations of effective noise-aware solutions are still lacking and require further investigation.

**Noise-aware training of the photonic generative model**

For our photonic GAN, we propose and experimentally validate two noise-aware training approaches, namely, the input-compensatory approach (IC-GAN) and the kernel weight-compensatory approach (WC-GAN), to improve the tolerance of the network to the effective weight setting noise $\Delta W^l_{ij}$. The IC-GAN approach inflates the STD of the random signal input from the experimental value of 0.2 V to 0.5 V during training. The WC-GAN approach adds $\Delta \Gamma^l_{ij}$ with 5% STD to the corresponding weight at each forward-propagation pass but performs noiseless gradient descent in the back-propagation pass (see Fig. 1b and Supplementary Information for the training procedure of these noise-aware training approaches). Fig. 3b and Fig. 3c show the experimentally generated images of handwritten "7" by the photonic GAN trained using both approaches. For a fair comparison, the random number inputs used for inferences are produced by the same optical RNG. Compared to the images generated by the noise-free trained GAN (NF-GAN) (Fig. 3a), the images generated using both noise-aware approaches display much clearer patterns with lower background noise, thus validating the noise tolerance of the IC-GAN and WC-GAN. Furthermore, we observe that the images generated by the WC-GAN (Fig. 3c) have richer handwritten-like features than those by the IC-GAN (Fig. 3b), with more diverse variations in styles. Therefore, we conclude that the WC-GAN is advantageous for practical implementation using non-ideal analog hardware.

**Discussion**

To quantitatively compare the GAN performance, we use the standard metric of Frechet inception distance (FID), which evaluates both the fidelity and diversity of the generated images by comparing the feature distribution in the generated images with images from the training dataset. The lower the FID score, the better performance of the GAN[51]. In Fig. 3d, the FIDs of the images generated by the NF-GAN[36,51,60], the IC-GAN, and the WC-GAN, respectively, are compared, assuming either ideal (FID$_{ideal}$) or noisy (FID$_{noisy}$) hardware (see Supplementary Information for detailed steps to calculate the FID). The FID$_{noisy}$ (hashed bars in Fig. 3d) is the lowest for the WC-GAN and the highest for the NF-GAN, consistent with the observation in Fig. 3a-c. The impact of hardware noise $\Delta$FID = FID$_{noisy}$ - FID$_{ideal}$ is plotted in Fig. 3e. The noise-aware WC-GAN and IC-GAN show two notable benefits. First, the FID$_{ideal}$ (solid bars in Fig. 3d) for the WC-GAN is lower than the NF-GAN (e.g., the software baseline[61]), indicating that introducing noises during training helps GAN learn better. Such a gain is absent in discriminative networks, where the inference accuracies of the noise-aware trained model cannot exceed the software baseline[29,30,33,34]. Second, surprisingly, the noise impact results (Fig. 3e) show that, unlike the NF-GAN, the WC-GAN and IC-GAN implemented on the photonic hardware with practical noise (hashed bars in Fig.3d) perform even better in inference than the noiseless hardware (solid bars in Fig. 3d). In contrast, a discriminative network's inference accuracy always decreases with more noisy hardware[37,38]. This surprising gain in performance suggests photonic neural networks' potential in generative models despite the inevitable optoelectronic noises and errors.

Optical computation in this work is performed at a low speed of 4000 operations per second (4 KOPS), limited by the use of the low-speed VOAs to encode data and the small-scale 2×2 tensor core. However, the state-of-the-art integrated photonic transmitters and photodetectors can drive the system at many 10s of Gbits/sec[64]. The size of the photonic core can be further scaled up to a much larger array. Assuming a moderate data rate of 10 Gbits/sec and 4 WDM channels, the computing density of a photonic tensor core can reach an upper-bound value of 25 TOPS/mm$^2$ (Tera-Operations per second per mm$^2$), significantly higher than that of the state-of-the-art digital electronics. To predict if the noise-aware approaches performance gain is scalable, in simulation, we train a larger-scaled GAN to generate images of all 10 number digits, using ideal or noise-aware approaches under various levels of writing errors. Fig. 4a shows the FID score of the results as a function of $\Delta\Gamma^I_{ij}$. Here, the curvature regularization approach (CR-GAN), which evolves from the WC-GAN, is used to improve the GAN robustness further (see the Supplementary Information for more details about the CR-GAN). The comparison shows that the CR-GAN performs better than the NF GAN at every error level. Note that under our present realistic noise level of 5% (Fig. 3g), the FID of CR-GAN is still below the software baseline, whereas the NF GAN's FID is higher than the baseline. For both approaches, with the increasing noise level, the FID first drops until reaching a minimum at ~2.5% noise and then increases. To explain this, we further examine the images generated by CR-GAN at three noise levels: 0%, 5%, and 10% in Fig. 4b-d. The comparison shows that the increasing hardware noise in GAN would improve the diversity (evaluated by the STD of the percentage of each number classes in the generated images[58], see Supplementary Information for more details) but at the same time reduce the fidelity of the generated images[58]. The trade-off results in a minimal FID at ~2.5% noise, as shown in Fig. 4a.

Throughout the full range of noise levels, the noise-aware approach consistently improves the GAN over the noiseless approach.

In conclusion, we demonstrate a photonic generative network based on phase-change photonics, which is used to form a GAN network and utilize the intrinsic noise sources in the photonic system. Unlike the previously demonstrated discriminative networks that suffer from the hardware noise, our experimental and simulation results show that the photonic generative network can not only tolerate but also benefit from a certain level of hardware noise after training by noise-aware training approaches. Our finding expands the current implementation of photonic neural networks to generative models[62], in which the inevitable and ubiquitous optoelectronic noises and errors can be mitigated and even leveraged in intelligent ways. We emphasize that the proposed noise-aware training approaches are generic and thus applicable to various types of optoelectronic neuromorphic computing hardware. The improved noise resilience of the model also implies their scalability in large-scale photonic neural networks with tightly co-integrated electronics and photonics.

**Materials and Methods**

**PMMC design and fabrication**

The PMMC consists of a phase-gradient metasurface made of GST thin film on silicon nitride waveguides. The metasurface is designed to convert the incident $TE_0$ mode into the $TE_1$ mode when GST is in the crystalline phase while maintaining the $TE_0$ mode when GST is in the amorphous phase. The PMMC is fabricated by depositing a 30nm thick GST film using a sputtering tool on an oxidized silicon substrate with 330 nm thick silicon nitride film. The GST film is then patterned into the metasurface using standard electron beam lithography (EBL) and inductively coupled plasma (ICP) etching processes. A 218 nm thick $Al_2O_3$ layer is deposited with atomic layer deposition to cap the GST conformally.

**Measurement setup**

The measurement set up to operate the photonic tensor core is shown in Fig. S2. The input optical signals are carried by four different wavelengths using four tunable CW lasers. The signal amplitudes are controlled by variable optical attenuators (VOA) with a 1 kHz operation speed. An additional control laser coupled with a 1×4 optical switch is used to optically program the kernel weight into each GST PMMC. The control pulses are generated with a 12 GHz electro-optical modulator and amplified by a low noise erbium-doped fiber amplifier (EDFA). The energy of each control pulse is further tuned using another VOA. The input signals and the control pulses are coupled into the photonic device via integrated grating couplers with a coupling efficiency of ~20%. The input signals propagate forward through each input channel while the control pulses propagate in the opposite direction through the $TE_1$ detection waveguides. The optical power in $TE_0$ mode is combined on-chip using integrated Y-junctions and detected. The optical power in the $TE_1$ mode is collected and combined off-chip. The mode power contrast is measured to give the MVM results.

**Acknowledgment:**

C.W., R.P. and M. L. acknowledge the funding support provided by the ONR MURI (Award No. N00014-17-1-2661) and the National Science Foundation (Award No. CCF-2105972). X.Y. and Y.C. acknowledge the funding support by NSF (Award No. 1955196) and ARO (Award No. W911NF-19-2-0107). H.Y., I.T. acknowledge the funding support provided by the ONR MURI (Award No. N00014-17-1-2661). Part of this work was conducted at the Washington Nanofabrication Facility / Molecular Analysis Facility, a National Nanotechnology Coordinated Infrastructure (NNCI) site at the University of Washington with partial support from the National Science Foundation via awards NNCI-1542101 and NNCI-2025489.

**Author Contributions:**

M.L. Y.C., and C.M. conceived the research and designed the experiments. C.M. designed and fabricated the photonic devices and performed the experiments with the assistance of R.P.. X.Y., C.M. design the neural network structure and performed simulation. H.Y. and I.T. developed and deposited the GeSbTe thin film. C.M., X.Y., M.L. analyzed the data. C.W., X.Y., and M.L. wrote the manuscript with contributions from all the authors. All authors discussed the results and commented on the manuscript.

**Data and Material Availability:**

All data needed to evaluate the conclusions in the paper are present in the paper and/or the Supplementary Materials. The codes of the numerical simulations are available at GitHub: https://github.com/cmwu2021/NoiseAware_Photonic_GAN/tree/main/Curvature_Regularization_GAN

**Competing Interests:**

The authors declare that they have no competing interests.

**FIGURES**

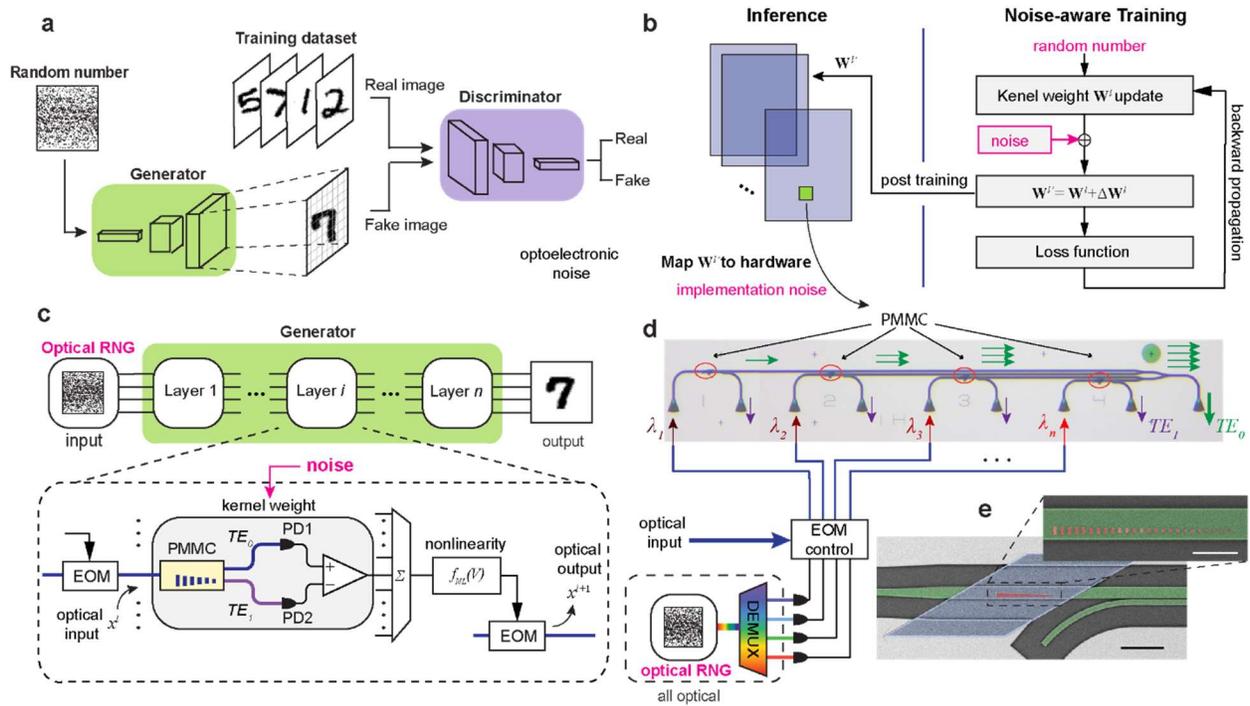

**Figure 1 Photonic GAN network with optoelectronic noises. a.** A GAN architecture is composed of two sub-network models, a generator and a discriminator. The generator competes with the discriminator during training and produces new instances after it is trained. **b.** The offline noise-aware training and inference processes flow of the generator. The process of mapping the trained weight to the hardware during implementation inevitably introduces optoelectronic noise. **c.** Decomposition of the generator into individual layers. In each layer, the input signals pass through the photonic tensor core and are converted to the electrical domain by photodetectors (PDs). After post-processing, the data is converted back into the optical domain and transferred to the next layer. **d.** Optical microscopic image of the photonic tensor core consisting of four input channels. The random noise is fed into the photonic tensor core through O/E and E/O conversion in our experiment. Potentially, the optical noise can be directly sent into the tensor core using WDM schemes. **e.** The detailed false-colored SEM image of the photonic tensor core. The $Si_3N_4$ waveguide, the GST metasurface, and the $Al_2O_3$ protection layer are colored green, red, and blue, respectively. Scale bar: 10 μm. Inset: the zoomed-in SEM image of the phase-gradient metasurface on the waveguide. Scale bar: 2 μm.
-12-

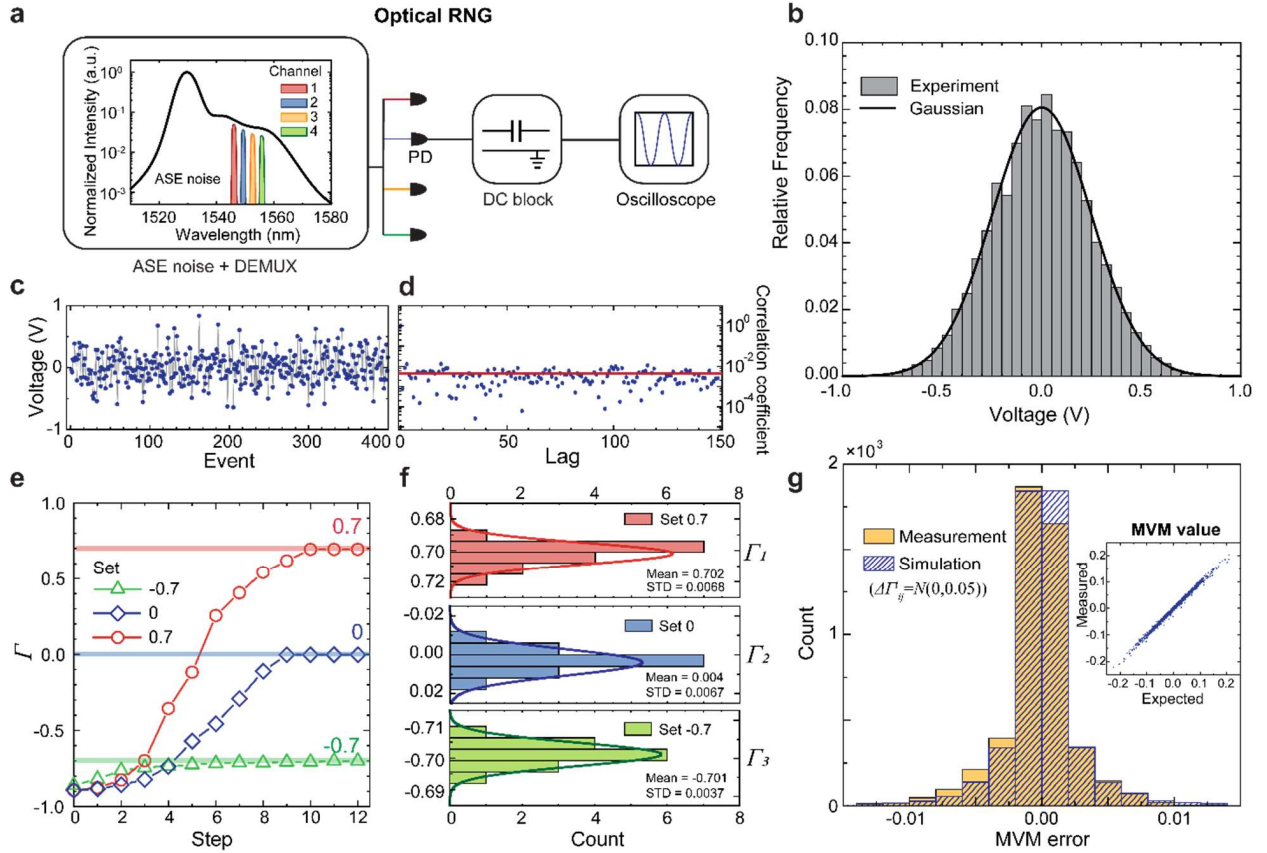

**Figure 2 Optical RNG and kernel programming errors. a.** Schematic of the optical RNG. The ASE noise is spectrally sliced into 4 wavelength channels using DEMUX and then detected with photodetectors. After a DC block, the random electrical signals are sampled by an oscilloscope. **b.** and **c.** Statistical histograms (**b**) and a representative trace (**c**) of the generated random numbers. The generated random number follows the Gaussian distribution. **d.** Correlation coefficient as a function of lag for the random number sequence. A random number sequence with length $N = 5×10^4$ has a correlation coefficient (blue dots) around the lower limit $1/\sqrt{N}$ (red line). **e.** Process of programming the mode contrast of a kernel element using optical pulses. The target $\Gamma$ values are -0.7, 0, and 0.7, respectively. **f.** Histogram of $\Gamma$ value distribution when the kernels are repeatedly set to be -0.7, 0, and 0.7, respectively. The STD for each setting is 0.37%, 0.67%, 0.68%, respectively. **g.** Histograms of the error distribution in the experimental measurement (solid) and the simulation (hashed) when assuming the $\Delta\Gamma^t_{ij}$ follow a Gaussian distribution with an STD of 5%. Inset: Measured MVM accuracy for 4900 MVM operations in the first layer of the network.

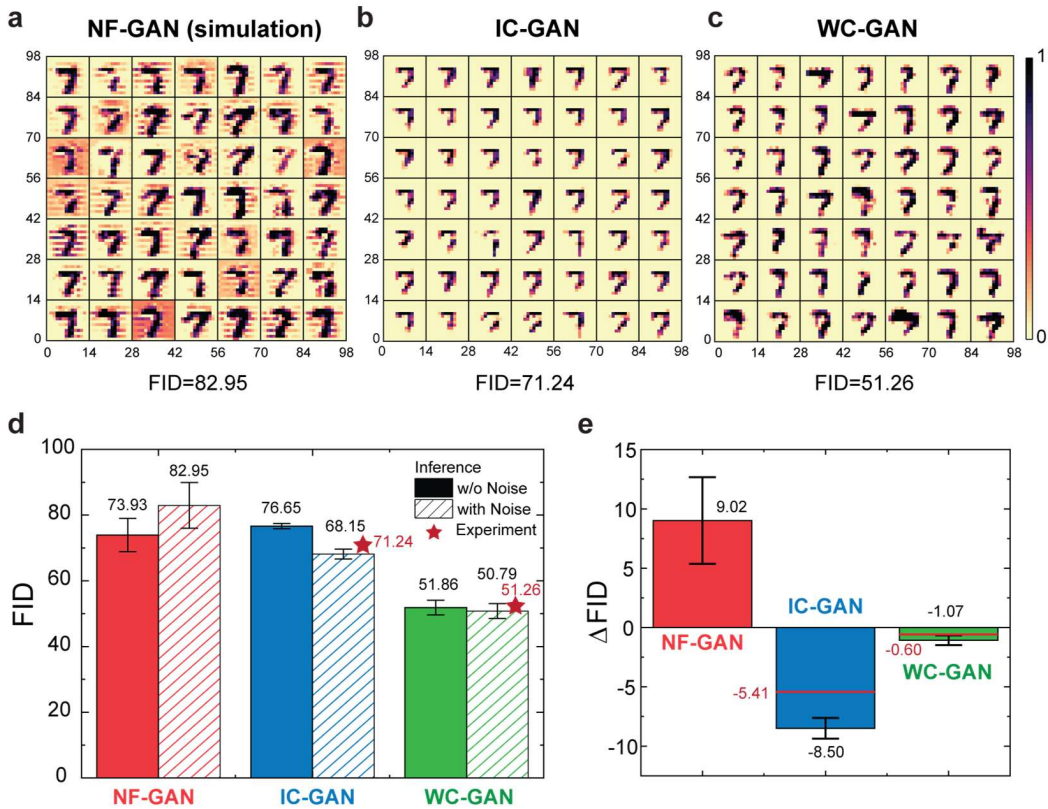

**Figure 3 Generating handwritten numbers with GAN. a-c:** 49 images (size: 14×14 pixels) generated by **(a)** NF-GAN, **(b)** IC-GAN, and **(c)** WC-GAN under effective kernel weight setting error (introduced by 5% Gaussian random error $\Delta \Gamma^t_{ij}$) and using random inputs~ $N(0,0.2)$ produced by the optical RNG. **(a)** is generated by simulation, **(b)** and **(c)** are from the experiments. **d.** The FIDs of the generated images, assuming the network is trained using various approaches and is implemented either on the ideal (solid bars) or noisy hardware (hashed bars). The FIDs obtained from the experimental results are labeled as stars. **e.** The difference of FID (ΔFID) in **(d)**. The ΔFIDs from the experimentally generated images are denoted by the red lines.

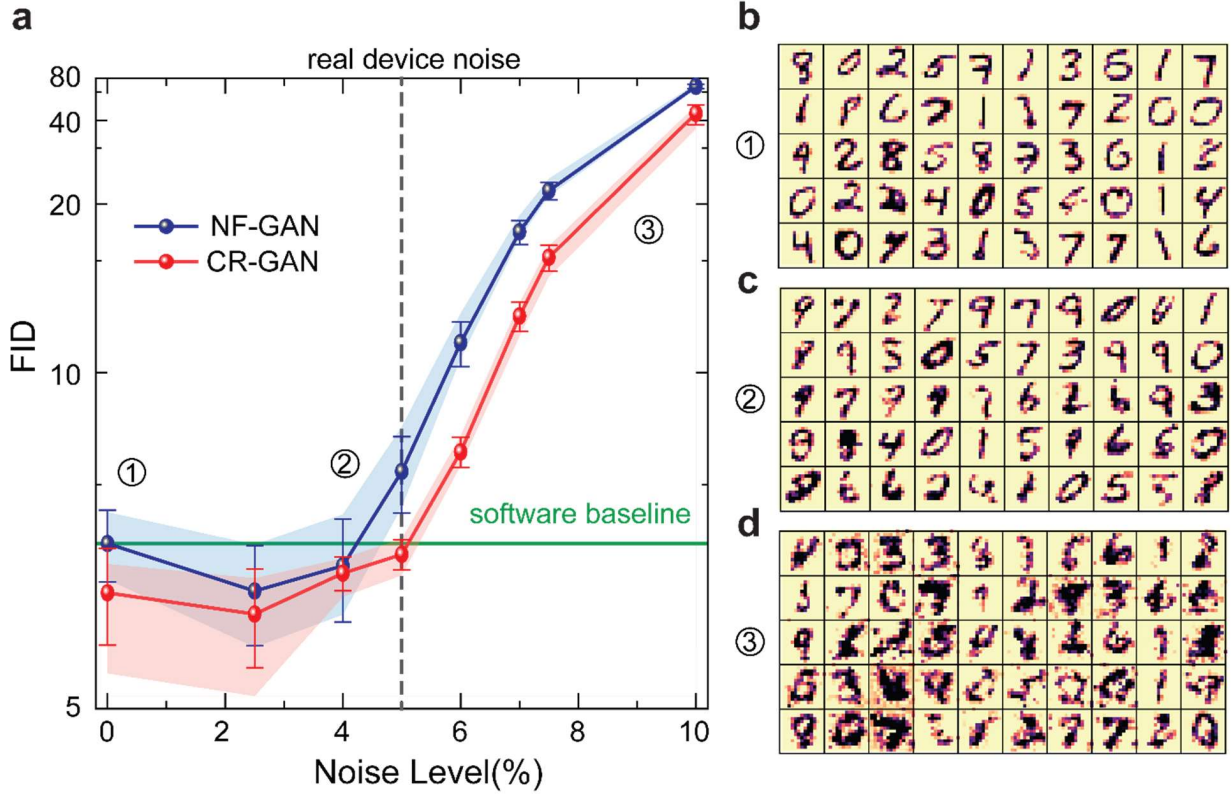

**Figure 4 Scalability of noise-aware training. a.** The FID of the generated images by the NF-GAN and the CR-GAN, respectively, under various effective mode contrast setting noise $\Delta\Gamma^l_{ij}$ with STD ranging from 0% to 10%. The shaded region indicates the range of FID over 5 individual tests. The FID is lower for CR-GAN at every noise level. At the measured noise level of 5% (black dashed line), the FID for CR-GAN is below the software baseline (solid green line) while the FID for the NF-GAN is above it. **b-d:** 50 images (size: 14×14) generated by CR-GAN assuming effective mode contrast setting noise of **(b)** 0%, **(c)** 5%, **(d)** 10%.